\documentclass[12pt]{article}
\usepackage{amssymb}
\usepackage{amsthm}
\usepackage{amsmath}
\usepackage{latexsym}
\usepackage{epsfig}
\usepackage{graphicx}
\usepackage{wasysym}
\usepackage{threeparttable}
\usepackage{natbib}
\usepackage{color}
\usepackage{caption}

\usepackage{hyperref}

\usepackage{comment}
\usepackage{float,subfigure}
\usepackage{setspace}

\def\boxit#1{\vbox{\hrule\hbox{\vrule\kern6pt
          \vbox{\kern6pt#1\kern6pt}\kern6pt\vrule}\hrule}}

\newcommand{\bbeta}{ \mbox{\boldmath $ \beta $} }

\newcommand{\balpha}{ \mbox{\boldmath $ \alpha $} }

\newcommand{\eps}{ \mbox{$\epsilon$}}

\newcommand{\btheta}{ \mbox{\boldmath $ \theta $} }

\newcommand{\bmu}{ \mbox{\boldmath $\mu$} }

\newcommand{\sig}{ \ensuremath{\sigma}}

\newcommand{\ind}{ \mbox{$\stackrel{\text{ind}}{\sim}$}}
\newcommand{\iid}{ \mbox{$\stackrel{\text{iid}}{\sim}$}}

\newcommand{\bzero}{\textbf{0}}

\newcommand{\bn}{ {\bf n} }

\newcommand{\bs}{ {\bf s} }

\newcommand{\bw}{ {\bf w} }
\newcommand{\bW}{ {\bf W} }
\newcommand{\bx}{ {\bf x} }
\newcommand{\bX}{ {\bf X} }

\newcommand{\bY}{ {\bf Y} }

\newcommand{\Cov}{\mbox{Cov}}

\newcommand{\given}{\,\vert\,}

\newcommand{\sS}{ {\cal S} }
\begin{document}
\thispagestyle{empty} \baselineskip=28pt

\begin{center}
{\LARGE{\bf Bayesian Marked Point Process Modeling for Generating Fully Synthetic Public Use Data with Point-Referenced Geography}}
\end{center}

\baselineskip=12pt

\vskip 2mm
\begin{center}
Harrison Quick\footnote{(\baselineskip=10pt to whom correspondence should be addressed) Department of Statistics, University of Missouri-Columbia, 146 Middlebush Hall, Columbia, MO 65211, quickh@missouri.edu},
Scott H. Holan\footnote{\baselineskip=10pt  Department of Statistics, University of Missouri-Columbia, 146 Middlebush Hall, Columbia, MO 65211-6100},
Christopher K. Wikle$^2$,
Jerome P. Reiter\footnote{\baselineskip=10pt  Department of Statistical Science, Box 90251, Duke University, Durham, North Carolina 27708-0251 }

\end{center}
%
%
%
%
\vskip 4mm

\begin{center}
\large{{\bf Abstract}}
\end{center}
Many data stewards collect confidential data that include fine geography. When sharing these data with others, data stewards strive to disseminate data that are informative for a wide range of spatial and non-spatial analyses while simultaneously protecting the confidentiality of data subjects' identities and attributes. Typically, data stewards meet this challenge by coarsening the resolution of the released geography and, as needed, perturbing the confidential attributes. When done with high intensity, these redaction strategies can result in released data with poor analytic quality. We propose an alternative dissemination approach based on fully synthetic data. We generate data using marked point process models {that} can maintain both the statistical properties and the spatial dependence structure of the confidential data. We illustrate the approach using data consisting of mortality records from Durham, North Carolina.
\baselineskip=12pt

%
%
%

\baselineskip=12pt
\par\vfill\noindent
\par\medskip\noindent
\clearpage\pagebreak\newpage \pagenumbering{arabic}
\baselineskip=24pt
\section{Introduction}\label{sec:intro}

{Many statistical agencies, research centers, and individual
researchers---henceforth all called agencies---collect confidential
data that they intend to share with others as public use files.
Many agencies are also obligated ethically, and often legally, to protect the
confidentiality of data subjects' identities and sensitive
attributes.  This can be particularly challenging for agencies seeking
to include fine levels of geography, e.g., street addresses or tax parcel
identifiers, in the public use files.  While detailed spatial information
offers enormous benefits for analysis, it also can enable ill-intentioned
users---henceforth called intruders---to easily identify individuals
in the file.

Because of these confidentiality risks, agencies typically alter
geographies and sensitive attributes before disseminating public use
files. Perhaps the most common redaction method is to aggregate geographies to
high levels like states (or not to release geography at all). Unfortunately,
aggregation sacrifices analyses that require finer
geographic detail and potentially creates  ecological fallacies \citep{wangreiter}.  Furthermore, when the file
includes other variables known to intruders like
demographic information, aggregation alone may not suffice to
protect confidentiality.  Another strategy is to move
each record's  observed location to another randomly drawn location, e.g., within some circle of radius $r$
centered at the original location. When large movements are needed to protect
confidentiality---as can be the case when released data include demographic and other
variables possibly known by intruders---inferences involving spatial relationships can be seriously
degraded \citep{rushton,guttman}.  Suppression and aggregation also
are commonly used to redact non-geographic
attributes \citep{willenborg, hundepool}, as are perturbative methods like data swapping
\citep{dalenius} and adding noise to values \citep{fuller93}.  When
applied with high intensity, these methods can result in files having
poor analytic quality without adequate confidentiality protection \citep{win07, holan2010, reit:drech:census}. 

An alternative to aggregation and perturbation is to release
multiply-imputed synthetic data, in which confidential values are
replaced with draws from statistical
models designed to capture important distributional features in the
collected data.  Synthetic data come in two
flavors.  Partially synthetic data comprise the original units
surveyed with some collected values replaced with multiple imputations
\citep{little93,kennickel:1997, abowd:2001, abowdwood04,reiter2003,
  reitermi,an:little:2007}, and fully synthetic data
comprise entirely simulated records  \citep{rubin93, reiter:2002, reiter:2002a,raghu:rubin:2001}.
In this article, we focus on fully synthetic data; see \citet{reiter:raghu:07} for a review of the
differences in the two flavors.  Fully synthetic data can offer low
disclosure risks as the released data cannot be meaningfully
matched to external databases, while allowing secondary analysts to make valid inferences for wide
classes of estimands via standard likelihood-based methods
\citep{raghu:rubin:2001, reitermulti}.



Thus far, synthetic data approaches have been used primarily for
data with no or highly aggregated geography  \citep[e.g.,][]{abowd06,
  hawalaacs, lbdisr} or with moderately aggregated geography like
block groups or areal regions
\citep[e.g.,][]{onthemap,burgettereiter,paiva}.  For the latter, the
basic idea is to aggregate from the point-level to discrete areal
units, estimate a model that predicts areal units from individuals'
attributes, and draw new areal units from the model as individuals'
synthetic locations.  These areal modeling approaches do not apply when the goal is
to release point-referenced geography, although \citet{paiva} make an
{\em ad hoc} suggestion to randomly assign each individual to a point
within its synthetic areal region.
One exception is the work of \citet{wangreiter}, who proposed that
agencies treat latitude and longitude just like other continuous variables,
approximating their  conditional distributions given non-synthesized
variables and releasing simulated locations by sampling from the models. They use
regression trees \citep{reiter2005} to approximate the conditional
distributions.  They also use trees to synthesize attributes conditional on latitude
and longitude, treating the geographies as predictors in the tree models.

{While the approach of \citet{wangreiter} is computationally efficient, it may not preserve local spatial dependence in the confidential data.
To do so more effectively, we propose to take advantage of models developed specifically for point patterns, in particular
models for marked point processes, to generate fully synthetic data with locations and attributes.
In marked point process modeling, there are two general approaches to modeling locations and attributes simultaneously.
Attributes (marks) can be modeled as conditional on the locations, where the locations are generated using a point process.
Typically, this approach combines standard geostatistical methods with point process theory to create a model where both the marks and the points are considered random \citep{diggletext}.
For example, \citet{rathbun}
create a space-time survival point process model for forest data where the birth of new trees is modeled using a point process,
 each tree's growth is modeled using a geostatistical model, and then the lifespan of each tree is modeled using a geostatistical survival model.
Alternatively, locations can be modeled as conditional on the marks, which is most sensible when the marks are categorical.  In essence, this results in a
collection of point processes---each with its own intensity surface---rather than a single point process.
\citet{liang}
used this approach to model rates of colon and rectal cancer in the greater Twin Cities (Minneapolis-St. Paul, MN) area.  {A similar approach has been used by
\citet{raj2011} in the context of
specifying knots in the modified predictive process of
\citet{banerjee2010}.}

In this article, we present an approach to generating fully synthetic, point-referenced data based on marked point process modeling.
We use a fully Bayesian hierarchical model that directly models data with exact geographic
locations, categorical marks (e.g., race, gender, education level,
cause of death), and non-categorical marks (subject's age).  We generate synthetic data { using} a three-step process, namely (i) generate synthetic versions of
the categorical attributes without considering locations, (ii) generate synthetic locations conditional on categorical attributes, and (iii) generate
synthetic versions of continuous attributes given the locations and categorical attributes.  We illustrate the approach on mortality records from Durham, North Carolina (NC).
We note that our approach differs from the that of \citet{ZhouEtAl2010}, who use spatial smoothing to mask non-geographic
attributes left at their original locations.

{The remainder of the article is organized as follows. In Section~\ref{sec:methods}, we describe the three-step method for modeling marked point processes.  In Section~\ref{sec:syn}, we describe how the parameter estimates from the marked point process model can be used to
generate fully synthetic datasets in a computationally efficient manner. 
In Section~\ref{sec:anal}, we illustrate the approach on publicly available mortality data from
Durham, North Carolina.  In particular, we demonstrate that the approach can preserve spatial structure in the original data and
offer high quality estimates of nonspatial regression coefficients. Finally, in Section~\ref{sec:disc}, we provide a concluding discussion.}

\section{Methodology for Modeling Marked Point Processes}\label{sec:methods}
The methodology we propose for modeling marked point processes is comprised of three main components: the categorical mark model, the point processes, and the non-categorical mark model.  These components build upon each other sequentially utilizing a hierarchical (conditional) framework.
\subsection{Model for categorical marks}\label{sec:catmark}
Suppose the data comprise $N$ individuals with $K$ combinations of
categorical marks.  Let $n_k$ denote the
number of individuals belonging to the $k$-th combination, where
$k=1,\ldots,K$.  A natural choice for modeling the vector $\bn =
(n_1,\ldots,n_K)'$ is the multinomial distribution,
\begin{equation}\label{eq:cat}
\bn\given N,\btheta \sim Mult(N,\btheta), 
\end{equation}
where $N = \sum_{k} n_k$ denotes the total sample size and $\btheta =
(\theta_1,\ldots,\theta_K)'$ denotes the vector of
probabilities. {It may be useful to put log-linear constraints on
$\btheta$ or to use mixtures of multinomial distributions
\citep{dunsonxing, sireiter13, manriquereiter14}, particularly when
$K$ is  large.}


\subsection{Point process model}\label{sec:pp}
To model locations, we take an approach similar to that of
\citet{liang}
and use a log-Gaussian Cox process for each categorical mark combination.  That is, for
a set of locations $\sS_k$  corresponding to the $k$-th combination, a spatial domain  $D$, and an intensity surface $\lambda_k(\cdot)$, we have
\begin{align}
LGCP\left\{\sS_k\given\lambda_k(\cdot),n_k\right\}&=\exp\left\{-\int_D \lambda_k(\bs) d\bs\right\} \prod_{i=1}^{n_k} \lambda_k(\bs_{i,k}),\label{eq:lgcp_int}
\end{align}
where $\log\lambda_k(\bs) = \bx_{\lambda}(\bs)'\bbeta_{\lambda\given
  k} + w_{\lambda\given k}(\bs)$. Here, $\bx_{\lambda}(\bs)$ denotes a
vector of {spatial predictors} (e.g., elevation, proximity to a body
of water, etc.) with the corresponding vector of regression
coefficients $\bbeta_{\lambda\given k}$, $w_{\lambda\given k}(\bs)$
denotes a random effect that induces correlation in the intensity
surface, and $\bs_{i,k} \in \sS_k$ for $i=1,\ldots,n_k$ and
$k=1,\ldots,K$.  In order to allow for correlation {between} intensity
surfaces, one option is to specify $\bw_{\lambda} =
(\bw_{\lambda\given 1}',\ldots,\bw_{\lambda\given K}')'$ such that
$\Cov(\bw_{\lambda}) =  \Psi_{\lambda} \otimes C_{\lambda}$, where
$\Psi_{\lambda}$ accounts for the covariance between surfaces and
$C_{\lambda}$ controls spatial association (say, using a Mat\'ern
correlation structure).

Due to the random structure of $\lambda_k(\cdot)$, the integral
in~(\ref{eq:lgcp_int}) is intractable and thus \eqref{eq:lgcp_int} cannot be computed in closed form.  Instead, we approximate the integral in~(\ref{eq:lgcp_int}) via numerical integration, leading to
\begin{align*}
LGCP\left\{\sS_k\given\lambda_k(\cdot),n_k\right\}&\approx \exp\left\{-\frac{|D|}{T} \sum_{i=1}^{n_{ni}} \lambda_k(\bs_{i,ni})\right\} \prod_{i=1}^{n_k} \lambda_k(\bs_{i,k}),
\end{align*}
where $\sS_{ni} = \{\bs_{i,ni}; i=1,\ldots,n_{ni}\}$ denotes the set of $n_{ni}$ points aligned using a grid for the numerical integration approximation.  The choice of $n_{ni}$ depends primarily on the spatial domain being analyzed, though larger values of $n_{ni}$ will result in a better approximation of the integral in~(\ref{eq:lgcp_int}), provided $\sS_{ni}$ covers $D$ evenly.

This increases the dimension of  $\bw_{\lambda}$ to $(n_k + n_{mc})$.  As $n_k$ may be large,
and as we may require $n_{ni}$ to be large { in order} to approximate
the integral in~(\ref{eq:lgcp_int}) accurately, this
can lead to a computational bottleneck.  As such, we use the predictive process of~\citet{banerjee2008} to reduce the dimension of $\bw_{\lambda}$ to a more manageable level.  In this framework, we define a set of $n^*$ knots, $\sS^* = \{\bs_1^*,\ldots,\bs_{n^*}^*\}$, $\bw_{\lambda}^* = (\bw_{\lambda\given 1}^{*'},\ldots,\bw^{*'}_{\lambda\given K})'$
and $\bw_{\lambda\given k}^* = (w_{\lambda\given k}^*(\bs_1^*),\ldots,w_{\lambda\given k}^*(\bs_{n^*}^*))'$ such that $\Cov(\bw_{\lambda}^* ) = \Psi_{\lambda} \otimes C_{\lambda}^*$. This results in
\begin{align}
LGCP\left\{\sS_k\given\widetilde{\lambda}_k(\cdot),n_k\right\}&\approx \exp\left\{-\frac{|D|}{T} \sum_{i=1}^{n_{mc}} \widetilde{\lambda}_k(\bs_{i,mc})\right\} \prod_{i=1}^{n_k} \widetilde{\lambda}_k(\bs_{i,k}), \label{eq:lgcp}
\end{align}
where $\log\widetilde{\lambda}_k(\bs) = \bx(\bs)'\bbeta_{\lambda\given k} + \widetilde{w}_{\lambda\given k}(\bs)$ with $\widetilde{w}_{\lambda\given k}(\bs) = C_{\lambda}(\bs)'(C_{\lambda}^*)^{-1}\bw_{\lambda\given k}^*$, where $C_{\lambda}(\bs)$ denotes the vector of spatial correlations between a location $\bs$ and the knot locations, $\sS^*$.  Details regarding the number and the placement of knots are provided in Section~\ref{sec:meth-comp}.

\subsection{Model for non-categorical marks}\label{sec:noncatmark}
{ The model for non-categorical marks proceeds using standard approaches from the geostastical literature; for further discussion see \citet{cressiewikle} and the references therein.  The exact model for the non-categorical marks is problem specific and agencies should use appropriate models for the non-categorical marks found in the data.  Here, we describe a model for a continuously varying mark based on a normal regression.  In Section~\ref{sec:anal}, where we model age as a non-categorical mark, we describe and use a truncated Poisson distribution.}


Let $Y_{k}(\bs_i)$ be the value of the non-categorical mark for
individual $i$ with categorical mark combination $k$.  For a
continuously varying $Y_{k}(\bs_i)$, a typical model is the regression
\begin{equation}\label{eq:Ybad}
Y_k(\bs) = \bx_Y(\bs)'\bbeta_{Y\given k} + w_{Y\given k}(\bs) + \epsilon_{Y\given k}(\bs),\;\text{where}\;\; \epsilon_{Y\given k}(\bs)\given\sig_k^2\; \iid N(0,\sig_{k}^2),
\end{equation}
where $\bx_Y(\bs)$ is a vector of spatially varying covariates with a corresponding vector of regression coefficients, $\bbeta_{Y\given k}$, and $w_{Y\given k}(\bs)$ is a random effect that induces correlation between the responses.  As before, we construct $\bw_Y$ with $\Cov(\bw_Y) = \Psi_Y \otimes C_{Y}$, where $\Psi_Y$ and $C_Y$ are defined similar to $\Psi_{\lambda}$ and $C_{\lambda}$, respectively. Here again, there may be computational issues pertaining to the dimension of $\bw_{Y\given k}$; thus, we consider the modified predictive process of
\citet{banerjee2010}.
This method is similar to the predictive process used for $\bw_{\lambda\given k}^*$, but here we have
\begin{equation}
\widetilde{w}_{Y\given k}(\bs)\given \bw_{Y\given k}^*,\phi_Y,\Psi_{Y} \ind N\left(C_{Y}(\bs)'(C_{Y}^*)^{-1}\bw_{Y\given k}^*, C_Y(\bs,\bs) - C_{Y}(\bs)'(C_{Y}^*)^{-1}C_{Y}(\bs)\right).\label{eq:modpp}
\end{equation}
The extra variability added from~(\ref{eq:modpp}) is necessary to correct for bias in the estimation of $\sig_k^2$ (see
\citet{finley} and \citet{banerjee2010}
for details).  Replacing $\widetilde{w}_{Y\given k}(\bs)$ for $w_{Y\given k}(\bs)$ in~(\ref{eq:Ybad}) yields
\begin{equation*} 
Y_k(\bs) = \bx_Y(\bs)'\bbeta_{Y\given k} + \widetilde{w}_{Y\given k}(\bs) + \epsilon_{Y\given k}(\bs).
\end{equation*}

\subsection{Linking the components}
We assume conditional
independence between the processes and parameters of the geographies
and the marks given the data, allowing
us to link the sub-models in Sections \ref{sec:catmark} --
\ref{sec:noncatmark} sequentially.
The resulting posterior distribution can be written as
\begin{align*}
[\btheta,
\bbeta_{\lambda}, \bw_{\lambda}^*,\phi_{\lambda},C_{\lambda},
\bbeta_Y&, \bw_Y^*, \widetilde{\bw}_Y, \{\sigma_k^2\},\phi_{Y},C_{Y} \given \bn,
\{\sS_k\},\bY]\notag\\ 
=& [\btheta \given \bn,
\{\sS_k\},\bY]\times [\bbeta_{\lambda},\bW_{\lambda},\phi_{\lambda},C_{\lambda}\given \btheta,\bn,
\{\sS_k\},\bY]\notag\\
&\times [\bbeta_Y, \bw_Y^*, \widetilde{\bw}_Y, \{\sigma_k^2\},\phi_{Y},C_{Y} \given \bbeta_{\lambda},\bw_{\lambda}^*,\phi_{\lambda},C_{\lambda},\btheta,\bn,
\{\sS_k\},\bY]\notag\\
=& [\btheta \given \bn]
\times [\bbeta_{\lambda},\bw_{\lambda}^*,\phi_{\lambda},C_{\lambda}\given \bn,
\{\sS_k\}]\\ 
& \times [\bbeta_Y, \bw_Y^*, \widetilde{\bw}_Y, \{\sigma_k^2\},\phi_{Y},C_{Y} \given \bn,
\{\sS_k\},\bY],\notag
\end{align*}
where, for two random variables $A$ and $B$, $[A\given B]$ denotes the conditional distribution of $A$ given $B$ and $[B]$
denotes the marginal distribution of $B$, $\bX$ denotes the matrix of covariate information, and $\bw_Y^*$, $\widetilde{\bw}_Y$, and $\bY$ are vectors analogous to $\bw_{\lambda}^*$.

Using the likelihood in \eqref{eq:cat}, for $\btheta$ we have
\begin{align*}
[\btheta \given \bn] \propto Mult(\bn\given N,\btheta) \times Dir(\btheta\given\balpha) 
\end{align*}
where $\balpha$ is a $K$-vector of probabilities such that $\sum \alpha_k = 1$. In practice, we  set $\alpha_k=1\slash K$ for all $k$ for a noninformative prior specification, though this choice can be adapted depending on the particular application.

Using the model in \eqref{eq:lgcp_int}, for
$(\bbeta_{\lambda},\bw_{\lambda}^*,\phi_{\lambda},C_{\lambda})$ we have
\begin{align*}
[\bbeta_{\lambda},\bw_{\lambda}^*,\phi_{\lambda},C_{\lambda}\given \bn,
\{\sS_k\}] \propto& \left(\prod_k LGCP(\sS_k\given \widetilde{\lambda}_k(\cdot),n_k)\right)
\times N(\bbeta_{\lambda}\given \bzero,\Sigma_{\beta,\lambda}) \\ 
&\times N(\bw_{\lambda}^*\given \bzero,\Psi_{\lambda} \otimes C_{\lambda}^*) \times Unif(\phi_{\lambda}\given a_{\phi},b_{\phi}) \times IW(C_{\lambda}^*\given \Gamma,\nu).\notag
\end{align*}
Here, we assign vague prior distributions for $\bbeta$ and
$C_{\lambda}^*$, and we let the hyperparameters for $\phi_{\lambda}$ depend on the
spatial range of the data. That said, $\phi_{\lambda}$ has been shown to be difficult to estimate in these types of models.
For instance, \citet{liang} opt
to fix $\phi_{\lambda}$ at a sensible value based on the maximum distance between knot points in the predictive process.  An alternative would be to compare the fits for a number of values of $\phi_{\lambda}$ and choose the value of $\phi_{\lambda}$ which provides the best fit based on some model selection criteria, turning an estimation problem into one of model selection.  

Using the likelihood in \eqref{eq:Ybad}, for $(\bbeta_Y, \bw_Y^*,
\widetilde{\bw}_Y, \{\sigma_k^2\},\phi_{Y},C_{Y})$ we have
\begin{align*}
[\bbeta_Y, \bw_Y^*, \widetilde{\bw}_Y, \{\sigma_k^2\},\phi_{Y},C_{Y} \given \bn,
\{\sS_k\},\bY] \propto& N(\bY\given \bmu,\Sigma_Y) \times N(\bbeta_{\lambda}\given \bzero,\Sigma_{\beta,Y}) \notag\\
& \times N(\bw_{Y}^*\given \bzero,\Psi_{Y} \otimes C_{Y}^*) \times \prod_k \pi(\sigma_k^2) \notag\\
& \times Unif(\phi_Y\given a_{\phi},b_{\phi}) \times IW(C_{Y}^*\given \Gamma,\nu) \notag\\
& \times [\widetilde{\bw}_Y\given \bw_Y^*,\Psi_{Y}, C_{Y}^*].
\end{align*}
Here, for $i=1,\ldots,n_k$ and $k=1,\ldots,K$, $\bmu$ is a mean vector
with elements $\mu_k(\bs_i) = \bx_Y(\bs_i)'\bbeta_{Y\given k} +
\widetilde{w}_{Y\given k}(\bs_i)$, and $\Sigma_Y$ is a block diagonal
matrix where the $k$-th block has $\sigma_k^2$ along the diagonal.  We
assign vague prior distributions for $\bbeta$ and $C_{Y}^*$, and let the hyperparameters for $\phi_Y$ depend on the spatial range of the data.
 Finally, $[\widetilde{\bw}_Y\given \bw_Y^*,\Psi_{Y}, C_{Y}^*]$ denotes the multivariate { normal} distribution from the modified
predictive process defined in~(\ref{eq:modpp}), and $\pi(\sigma_k^2)$ denotes an improper uniform prior for $\sigma_k$ on the interval $(0,\infty)$, following the recommendation of
\citet{gelman2006}. 


\subsection{Computational details}\label{sec:meth-comp}
A benefit of the decomposable nature of this model is that
the model can be estimated in {parallel}.  That is, the MCMC algorithm
can be run for each sub-model independently of the others, which saves
computing time and resources.
Further  parallelization and hence computational savings can be
realized by estimating the models independently for each categorical
mark combination,  though
this would require either fixing the range parameter(s),
$\phi_{\lambda}$ and/or $\phi_Y$, at sensible values or allowing each
combination to have its own range parameter, e.g.,
$\phi_{\lambda,k}$.  These actions can limit borrowing of
strength across {combinations}; thus, we recommend this strategy only when all $n_k$ are sufficiently large.
In the NC mortality data synthesis, we fix both $\phi_{\lambda}$ and $\phi_Y$ to
facilitate faster computation. We recommend assessing
sensitivity by refitting the model with different choices of these
parameters.

As in any knot-based method, the number and placement of knots are
important decisions.  Regarding the placement of knots, it is quite
common to simply place knots by overlaying a grid of the spatial
region \citep[e.g.][]{liang}.
Others suggest the use of more sophisticated methods, such as space filling designs \citep{nychka98}.
In the case of the modified predictive process,
\citet{raj2011}
used a point process model to adaptively select knots throughout the course of their MCMC algorithm.  The authors found that an adaptive knot selection algorithm with only 25 knots performed as well as a non-adaptive algorithm with 81 knots aligned on a grid with respect to model fit and was on par with 225 knots with respect to prediction, despite requiring only 1/3 of the computing time.  

Here, we offer a compromise by splitting the $n^*$ knots into two
subsets.  First, we place $n_g^*$ knots  uniformly using a grid over $D$, ensuring that we learn about the entire intensity surface.
For the remaining $n_{pp}^*=n^*-n_g^*$ knots, we first fit a general (e.g., unmarked) point process for the data to find the overall intensity surface; this can be done using the approach described in Section~\ref{sec:pp} using $K=1$.  After determining where the data are more heavily concentrated, we draw $n_{pp}^*$ knot locations using this overall intensity surface.  Unlike \citet{raj2011},
however, we do  not resample knot locations at each iteration of the MCMC, thereby avoiding additional computational burden.  With this approach, our goal is to achieve the benefits of dense knot placement (increased predictive performance) and learn about the entire surface by filling the space, while preserving the computational benefits of having fixed knot locations.
{In general, we recommend choosing $n_g^*=n_{pp}^*$ such that $n^*<n_k$ for all $k$}.  Furthermore, based on the existing predictive process literature, we choose values of $n^*$ near 100 --- aiming to strike a balance between predictive performance and computational burden --- though in practice this will depend on the particular dataset.

\section{Generating Fully Synthetic Data}\label{sec:syn}
Suppose we seek to generate $L$ fully synthetic datasets of size
$N$.
We  assign categorical marks to each of the $N$ samples by drawing
from their posterior predictive distribution.  Under the
multinomial-Dirichlet model, we can create synthetic mark combinations
for all $N$ records in one step.  Let $\bn^{\dagger(\ell)} =
\{n_k^{\dagger(\ell)}: k = 1, \dots, K\}$ be the number of cases at
each categorical mark combination in the $\ell$-th synthetic dataset. We
sample $\bn^{\dagger(\ell)}$ using
\begin{align*}
\bn^{\dagger(\ell)}\given \btheta^{(\ell)} \sim Mult\left(N,\btheta^{(\ell)}\right),
\end{align*}
where $\btheta^{(\ell)}$ is an approximately independent draw from the posterior
distribution of $\btheta$.

{For each record, we next draw a synthetic location from the intensity
surface corresponding to its synthetic categorical mark combination. 
To do so conveniently, we first choose a large number of candidate locations; these serve as the sample space of the synthetic locations.
For each of our $N_s$ candidate locations, $\bs^{\dagger}$, we compute $\widetilde{\lambda}_k^{(\ell)}(\bs^{\dagger})$ from the $\ell$-th sample from the posterior predictive distribution.  Given our estimated intensity surface, $\widetilde{\lambda}_k^{(\ell)}(\cdot)$, we can sample from
\begin{align*}
\sS_k^{\dagger(\ell)}\given\widetilde{\lambda}_k^{(\ell)}(\cdot),n_k^{\dagger(\ell)} \sim LGCP\left\{\widetilde{\lambda}_k^{(\ell)}(\cdot),n_k^{\dagger(\ell)}\right\}
\end{align*}
by letting
\begin{equation*}
P_k\left(\bs_i^{\dagger(\ell)} = \bs_j\right) = \widetilde{\lambda}_k^{(\ell)}(\bs_j) \slash \sum_{j=1}^{N_s} \widetilde{\lambda}_k^{(\ell)}(\bs_j) \;\text{for}\; i=1,\ldots,n_k^{\dagger(\ell)}, k=1,\ldots,K.
\end{equation*}

When available, the candidate pool can include all actual locations (e.g., residential addresses) in $D$.  Otherwise, the pool can be drawn uniformly across $D$ or placed using a fine grid.  In the latter case, when the size of the candidate pool is not substantially larger than $N$, or when certain areas are particularly densely populated, we recommend sampling the $n_k^{\dagger(\ell)}$ locations from the pool with
replacement.
While sampling with replacement can generate  multiple observations at
the same location, this is preferred to drawing unlikely locations
(with respect to  $\bn^{\dagger(\ell)}$ and the estimated intensity surface)
solely due to an inadequate number of candidate locations.
As such, we recommend sampling with replacement using at least 2,500
candidate locations (e.g., a $50\times50$ grid). Of course, the size
of the pool  should depend on $D$ and the spatial
resolution desired.

Once the synthetic individuals have been assigned categorical marks
and locations, we next generate their non-categorical marks from their
posterior predictive distributions.  For example, when
$Y_{k}(\bs_i)$ is assumed to follow \eqref{sec:noncatmark}, we sample
each synthetic  $Y_k^{\dagger(\ell)}(\bs_i^{\dagger(\ell)})$ from
\begin{align*}
Y_k^{\dagger(\ell)}(\bs_i^{\dagger(\ell)})\given \mu_k^{(\ell)}(\bs_i^{\dagger(\ell)}),\sigma_k^{2(\ell)} \sim N\!\left(\mu_k^{(\ell)}(\bs_i^{\dagger(\ell)}),\sigma_k^{2(\ell)}\right), 
\end{align*}
where $(\mu_k^{(\ell)}, \sigma_k^{2(\ell)})$ is an approximately independent draw from the posterior
distribution of $(\mu_k, \sigma_k^{2})$, for $k=1, \dots, K$.
When $Y_{k}(\bs_i)$ is assumed to follow other models, we sample from
the corresponding posterior predictive distribution, and similarly for the case of multiple non-categorical marks.}

\subsection{Evaluating data utility}
To evaluate the utility of the fully synthetic data, we recommend comparing the distribution of synthetic and original geographies as well as the results of representative statistical
analyses on the synthetic and original data.  For geographies, we compare the $K$ function of the synthetic geographies to that of the confidential data.  The $K$ function
\citep{bartlett64,ripley76}
is a measure of spatial dependence such that, for a given radius, $h$, $K(h)$ is the expected number of events within distance $h$ of an arbitrary event.  An estimate of $K(h)$ can be obtained by computing
\begin{equation}\label{eq:kfun}
\widehat{K}(h) = \frac{|D_s|}{N} \underset{i\ne j}{\sum_{i=1}^N \sum_{j=1}^N} I(||\bs_i-\bs_j||\le h)\slash N.
\end{equation}
where $|D_s|$ is the area of spatial domain.  As discussed in
\citet[pp.\ 210--212]{cressiewikle},
this estimator is biased in the presence of edge effects; a discussion of edge-corrected estimators can be found in
\citet[pp.\ 615--618]{cressietext}.
Using~(\ref{eq:kfun}), we can then compute an estimate of the $L$ function,
\begin{equation}\label{eq:lfun}
\widehat{L}(h) = \sqrt{\widehat{K}(h)\slash\pi} - h,
\end{equation}
where positive values of $\widehat{L}(h)$ indicate spatial clustering. For our purposes, we compute~(\ref{eq:lfun}) for a range of values of $h$ in each of the synthetic datasets and obtain the average over $h$, denoted $\widehat{K}^{\dagger}(h)$, and then compare the resulting curve (and its pointwise empirical 95\% CI) to that obtained using the original data.  

With respect to analyses of the marks, we use the methods in \citet{reiter2003}
 to determine point and interval estimates from the synthetic data.  These inferential methods---developed for
partially synthetic data---are also appropriate for fully synthetic data when
(i) the original data can be analyzed as a simple random sample, and (ii)
the number of synthetic observations equals the number of original observations.  When this is not the case, we
recommend using the methods of \citet{sireiter} or \citet{raghu:rubin:2001}.


\subsection{Evaluating disclosure risks}\label{sec:risk}

Intruders cannot meaningfully match fully synthetic records to external files; however, this does not mean that the synthetic data are risk free.  In particular, the
synthetic data can be subject to inferential disclosure risks \citep{hundepool}, i.e., intruders can use the synthetic data to estimate confidential data values with high accuracy.
In this section, we describe two inferential disclosure risk measures,
one for estimating
spatial locations given marks and one for estimating attributes given locations.

Before presenting the measures, we define two terms: \emph{spatially
  close} and \emph{similar attributes}.  Two locations are said to be
spatially close if the distance between them is less than
$\epsilon_s$, where $\epsilon_s$ is defined by the agency.
Two individuals are said to have similar attributes if they belong to
the same mark categories and their non-categorical marks are within
$\epsilon_a$ of each other, where again $\epsilon_a$ is defined by the agency.
The definition for similar attributes assumes a single non-categorical
mark, but it could be extended to multiple non-categorical marks.  For
an individual at location $\bs_0$ with categorical mark combination
$k$ and non-categorical mark $Y_{k}(\bs_0)$, we denote the properties
of being spatially close and having similar attributes as $\sim\bs_0$
and $\sim Y_{k}(\bs_0)$, respectively.

In the first scenario, we assume that an intruder knows someone is at
a particular location $\bs_0$ and seeks to learn that individual's attributes.
 In this setting, we consider the inferential disclosure risk to be high if a large percentage of synthetic individuals who are spatially close to $\bs_0$ have similar attributes
to an individual in the confidential dataset at location $\bs_0$.  To
assess this,  we estimate
$
P(\sim Y_{k}(\bs_0) \given \sim \bs_0) 
$ 
by computing
\begin{align}
p_s^{\dagger(\ell)}(Y_k(\bs_0),\bs_0) = \frac{\sum_i I\{Y_k^{\dagger(\ell)}(\bs_i^{\dagger(\ell)}) \sim Y_k(\bs_0) \given \bs_i^{\dagger(\ell)} \sim \bs_0\}}{\sum_i I\{\bs_i^{\dagger(\ell)} \sim \bs_0\}} \label{eq:ps}
\end{align}
for $\ell=1,\ldots,L$.  Uncertainty estimates for {this type of risk --- henceforth referred to as ``Type S Risk'' ---} can be found using the quantiles of $\left\{p_s^{\dagger(\ell)}(Y_k(\bs_0),\bs_0)\right\}_{\ell=1}^{L}$.

In the second scenario, we assume than an intruder knows an
individual's attributes and wishes to learn the location.  Here, the
inferential disclosure risk for an individual at location $\bs_0$ is high if a large
percentage of synthetic individuals with similar attributes are
spatially close to $\bs_0$.  To assess this, we estimate
$
P(\sim \bs_0 \given \sim Y_k(\bs_0))
$ 
by computing
\begin{align}
p_a^{\dagger(\ell)}(Y_k(\bs_0),\bs_0) = \frac{\sum_i I\{\bs_i^{\dagger(\ell)} \sim \bs_0 \given Y_k^{\dagger(\ell)}(\bs_i^{\dagger(\ell)}) \sim Y_k(\bs_0)\}}
{\sum_i I\{Y_k^{\dagger(\ell)}(\bs_i^{\dagger(\ell)}) \sim Y_k(\bs_0)\}} \label{eq:pa}
\end{align}
for $\ell=1,\ldots,L$.  As before, uncertainty estimates for {this type of risk --- henceforth ``Type A Risk'' --- }can be found using the quantiles of $\left\{p_a^{\dagger(\ell)}(Y_k(\bs_0),\bs_0)\right\}_{\ell=1}^{L}$.

As written, both~(\ref{eq:ps}) and~(\ref{eq:pa}) assume that each synthetic dataset contains at least one observation that is spatially close and at
least one observation that has similar attributes.  This may not always be the case for any given
 $\eps_s$ or $\eps_a$, particularly for outlying observations (either spatially or with respect to the attributes).  When this occurs, one option
 is to choose sufficiently large values for these thresholds to ensure that neither~(\ref{eq:ps}) and~(\ref{eq:pa}) has a zero in the denominator.
 An alternative is to use only  the synthetic datasets with non-zero denominators; for instance,
 if only 40 synthetic datasets have spatially close individuals for a given $\bs_0$, base the estimate for $P(\sim Y_{k}(\bs_0) \given \sim \bs_0)$
only on these 40 datasets.

\section{North Carolina Mortality Data Example}\label{sec:anal}
We now apply the methods presented in Section~\ref{sec:methods} to
create fully synthetic datasets for a database on causes of mortality
in Durham, NC in the year 2002. The data comprise $N=6294$
records that include the precise latitude and longitude of each
individual's residence at time of death (these have been scaled to the unit
square).  Each record also includes the individual's age (between 16 and 98), race (black or white),
sex, level of education, and cause of death.  We categorize the level of education into less than high
school, high school, and some college.  We dichotomize cause of death into an indicator for causes due to cancer or a failure of the immune system versus all other causes.  Similar data were used by \citet{wangreiter} and \citet{paiva}.

The cross-tabulation of race, sex, education, and the cause of death
indicator results in $K=24$ combinations, which we treat as the
categorical marks.  The values of $n_k$ range from 79 to 525, so that
we have sufficient sample size to estimate the intensity surface for each combination
separately.   We treat age as a
non-categorical mark.  Since age is integer-valued and has restricted support, we
model it using the truncated Poisson distribution,
\begin{equation*}
Y_k(\bs_i)\given\beta_{Y\given k},\widetilde{w}_{Y\given k}(\bs_i) \sim TrunPois\left(\exp\left\{\beta_{Y\given k}+\widetilde{w}_{Y\given k}(\bs_i) \right\},[16,98]\right),
\end{equation*}
where $\widetilde{w}_{Y\given k}(\bs)=
C_{Y}(\bs)'(C_{Y}^*)^{-1}\bw_{Y|k}^*$ is from a predictive process
akin to that used in~(\ref{eq:lgcp}).
We use an intercept-only model as we do not have spatial covariates.
We use a flat prior for each $\beta_{Y\given k}$ and conventional
prior specifications for other parameters, so that the sub-model for age is
\begin{align*}
[\bbeta_Y, \bW_Y^*, \phi_{Y},C_{Y} \given \bn,
\{\sS_k\},\bY] \propto&
\prod_k \prod_i TrunPois\left(Y_k(\bs_i)\given\exp\!\left\{\beta_{Y\given k}+\widetilde{w}_{Y\given k}(\bs) \right\}\!,[16,98]\right)\notag\\
& \times Unif(\phi_Y\given a_{\phi},b_{\phi}) \times IW(C_{Y}^*\given \Gamma,\nu)
\times N(\bW_{Y}^*\given \bzero,\Psi_{Y} \otimes C_{Y}^*). 
\end{align*}
Here again, we opt to fix $\phi_{Y}$ so that the effective range is half the maximum
distance between any two points in our original dataset.
To estimate the intensity and age surfaces, we use $n^*=72$ knots, {allowing $n_g^* = n_{pp}^*$ while also ensuring $n^* < n_k$ for all $k$}.

We generate $L=100$ synthetic datasets, each comprised of $N$ individuals.  To do so, we follow
the approach described  in Section~\ref{sec:syn}, replacing the
normal model with the truncated Poisson model. For the pool of candidate synthetic locations, we sample locations
from a $50\times50$ grid, as we do not have a list of
all possible locations in Durham.
We generate synthetic ages from the appropriate posterior predictive distribution, which given parameter draws takes the form
\begin{equation*}
Y_k^{\dagger(\ell)}(\bs_i^{\dagger(\ell)})\given\beta_{Y\given k}^{(\ell)},\widetilde{w}_{Y\given k}^{(\ell)}(\bs_i^{\dagger(\ell)}) \sim
TrunPois\left(\exp\left\{\beta_{Y\given k}^{(\ell)}+\widetilde{w}_{Y\given k}^{(\ell)}(\bs_i^{\dagger(\ell)})\right\},[16,98]\right).
\end{equation*}

To evaluate the utility of the synthetic data, we first compare the
synthetic and original locations with the $L$ function.
Figure~\ref{fig:mort_int} displays the comparison for white males with
less than a high school education whose cause of death was not due to
cancer or some failure of the immune system; {similar findings hold for the remaining 23 groups}.
For $h \geq .04$, $\widehat{L}^\dagger(h)$ from the synthetic data is similar to
$\widehat{L}(h)$ from the confidential data, indicating that the synthetic data
accurately reflect the degree of clustering in the real data for these values of $h$.
When $h < .04$, $\widehat{L}^\dagger(h)$ slightly underestimates
$\widehat{L}(h)$ from the confidential data.  This
bias may be due to the discontinuity in $\widehat{K}^\dagger(h)$
resulting from generating data on a $50\times50$ grid.

\begin{figure}[t] 
    \begin{center}
        \subfigure[True Locations]{\includegraphics[width=3in, height=3in]{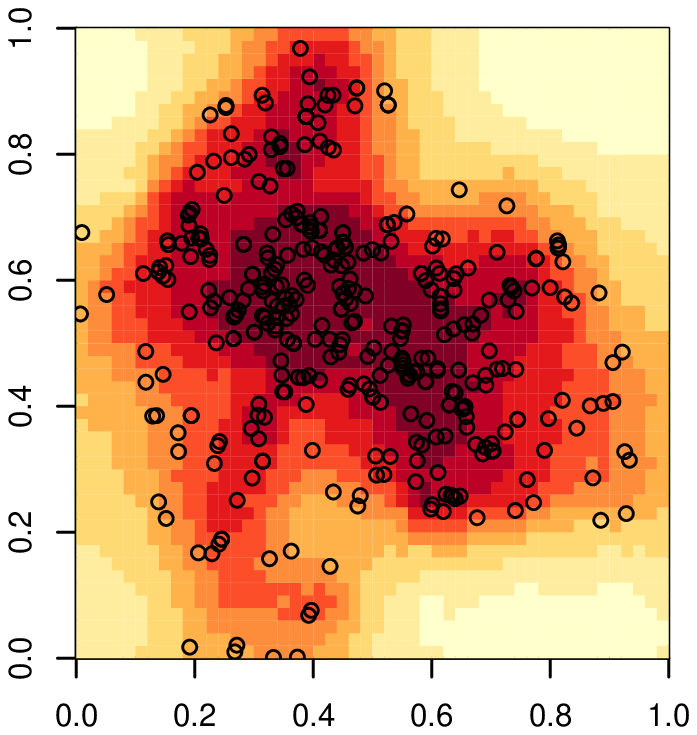}\label{fig:true}}
        \subfigure[Synthetic Locations]{\includegraphics[width=3in, height=3in]{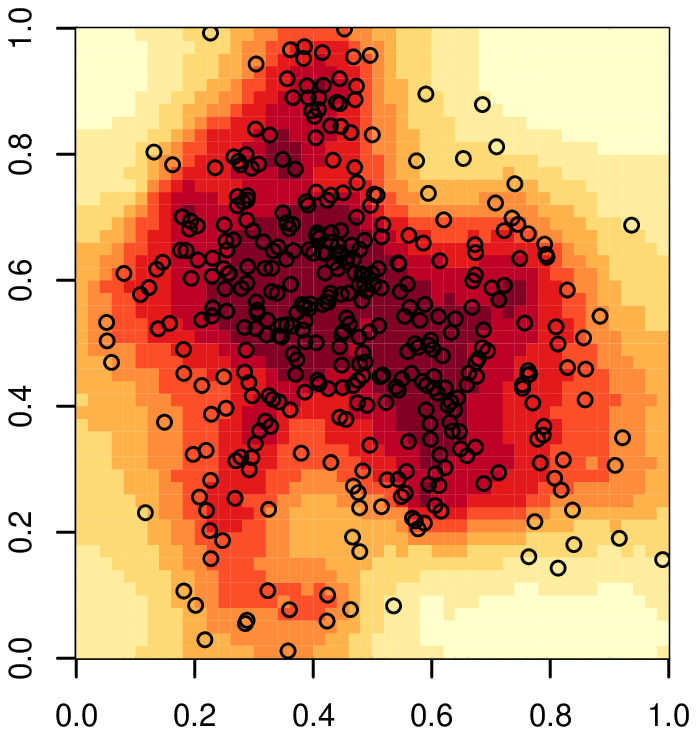}\label{fig:syn}}\\
        \subfigure[$L$ Function]{\includegraphics[width=3in, height=3in]{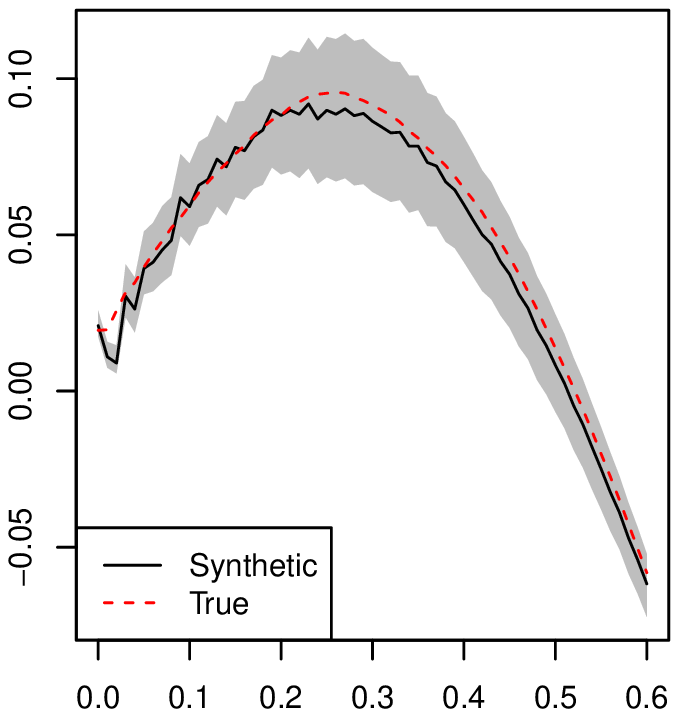}}
    \end{center}

   \caption{Selected figures for white males with less than a high school education whose cause of death was not due to cancer or some failure of the immune system from the Durham, NC mortality data.  Panel~(a) displays locations of the observations from the true dataset while panel~(b) displays the geographies from one of our synthetic datasets, both overlayed on our estimated intensity surface where off-white denotes low intensity and red denotes high.  Panel~(c) displays $\widehat{L}(h)$ over a range of values of $h$ for our true (dashed red line) and synthetic data (black line), along with the pointwise empirical 95\% CI from the synthetic data (gray band).}
   \label{fig:mort_int}
\end{figure}

We next evaluate two analyses involving the marks.  First,
we fit a Poisson regression  of age on the set of $K$ indicator variables for each categorical mark combination, i.e.,
\begin{equation*}
\text{Age}_i \sim Pois(\lambda_i) \;\;\text{where}\;\;\log(\lambda_i) = \sum_{j=1}^{24} \beta_j I(\text{Group}_i = j),
\end{equation*}
where $\text{Group}_i$ is an integer between 1 and 24 denoting which combination of race, gender, education level, and cause of death the individual belongs to.
Figure~\ref{fig:pois} displays the estimated coefficients from this regression for both the synthetic and confidential data.  Here,
we use the methods from \citet{reiter2003} {to form} inferences from the synthetic data.  The inferences from the
synthetic data are quite similar to those based on the confidential data.

Second, we fit a logistic regression of cause of death on race, sex, and education level and their interactions; i.e., we let $\text{Group2}_i$ be an integer between 1 and 12 denoting which combination of race, gender, and education level each individual belongs to and assume
\begin{equation*}
\mbox{logit}(\pi_i)=\log\{\pi_i/(1-\pi_i)\} = \sum_{j=1}^{12} \beta_j I(\text{Group2}_i = j).
\end{equation*}
Figure~\ref{fig:log} displays the estimated coefficients from the synthetic and confidential data.  Once again, the
synthetic data inferences are similar to the confidential data inferences.
Of note, the cause of death in females without a high school education --- in both white and black
populations (groups 7 and 10 in Figure~\ref{fig:log}, respectively) --- is significantly less likely to be
cancer-related than it is for the rest of the population.

\begin{figure}[t] 
    \begin{center}
        \subfigure[Poisson Regression]{\includegraphics[width=3in, height=3in]{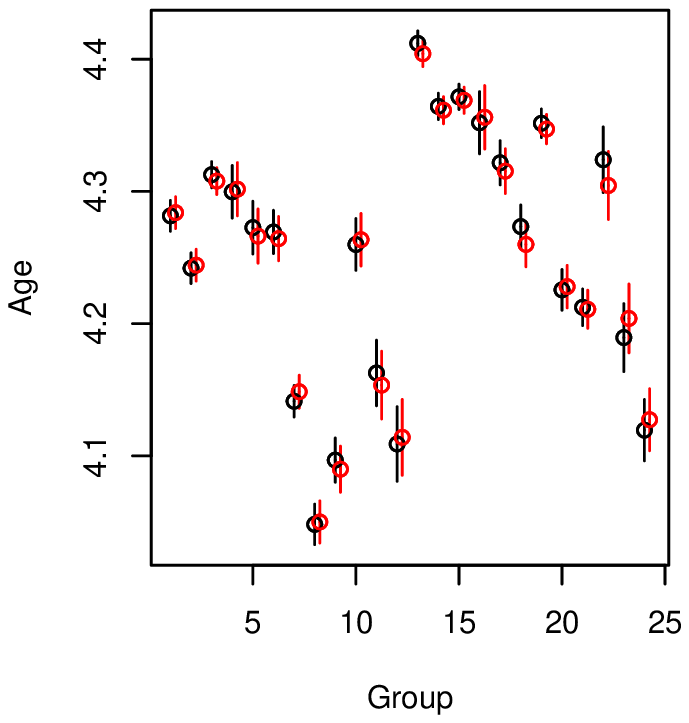}\label{fig:pois}}
        \subfigure[Logistic Regression]{\includegraphics[width=3in, height=3in]{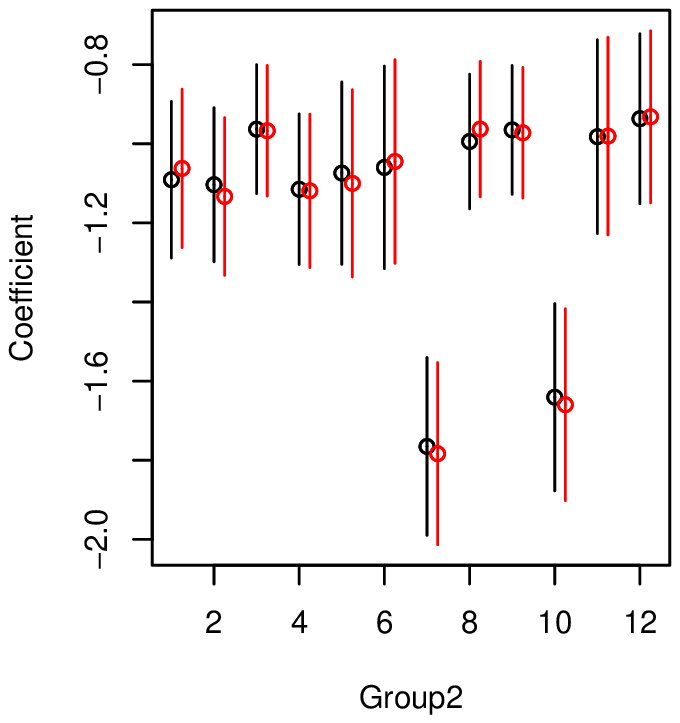}\label{fig:log}}\\
        \subfigure[Risk]{\includegraphics[width=3in, height=3in]{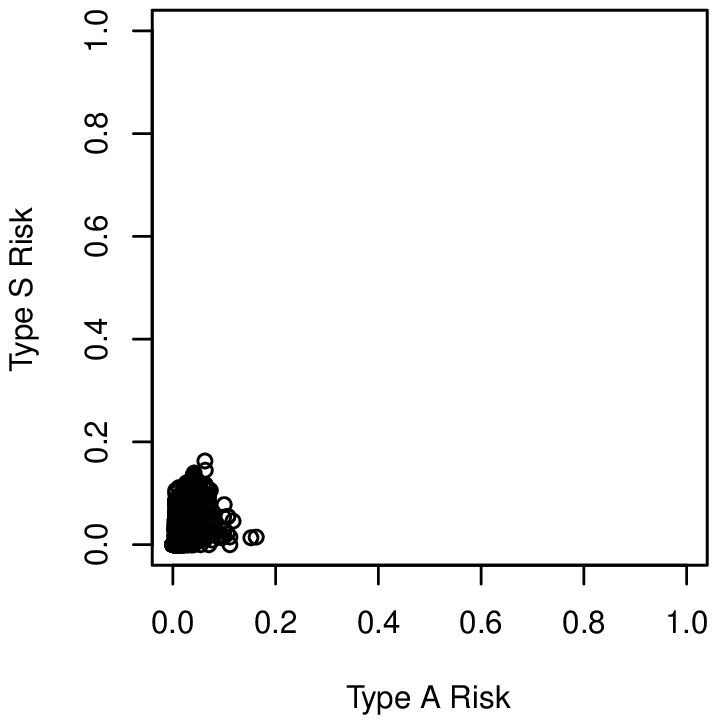}\label{fig:mort_risk}}

    \end{center}
   \caption{Assessments of data utility and risk.  Panels~(a) and~(b) compare parameter estimates from a Poisson regression on age and a logistic regression on the cause of death.  Circles denote mean estimates while bars denote the 95\% CI.  Estimates from the real data are displayed in black (the leftmost interval in each pair), while estimates obtained from the synthetic data are displayed in red (the rightmost interval in each pair).  Panel~(c) plots the estimated Type S Risk versus the estimate Type A Risk. Estimates shown here are the medians of the values of~(\ref{eq:ps}) and~(\ref{eq:pa}) over our 100 synthetic datasets.}
   \label{fig:regression}
\end{figure}

These evaluations of data utility suggest that the synthesizer can generate data with useful analytic properties.
{Of course, it is prudent for agencies to evaluate the quality of synthetic data on a much wider class of representative analyses;
see \citet{2stagesyn} for further discussion of this issue.}
We also require the synthesizer to generate synthetic data with reasonably low inferential disclosure risks.  To evaluate this, we use
the risk measures in~(\ref{eq:ps}) and~(\ref{eq:pa}).  As shown in Figure~\ref{fig:mort_risk}, all 6294 individuals in the confidential
data have estimated levels of Type S and Type A risk less than 0.20 when using $\eps_s = 0.02$ {(roughly $1\slash4$ of a mile)}
and $\eps_a = 5$ years.  Furthermore, many of the individuals who we identify as being at the highest risk are
simply those from the most frequently observed sub-populations in the data.  Given the nature of these data, it is quite
possible that a number of these individuals resided in assisted living facilities in which a number of
elderly persons with similar racial and socioeconomic backgrounds live.

\section{Discussion}\label{sec:disc}

In addition to providing fully synthetic data with potentially high utility, there are several other benefits to the hierarchical marked point process approach.  
This approach allows one to model the various intensity surfaces and the response surface(s) in parallel
(whether or not one fixes the spatial range parameter, $\phi$), {yielding substantial benefits both computationally and in terms of model simplicity}.
In contrast, modeling a single intensity surface and jointly modeling a large number of responses
would result in a substantially higher computational burden.  In fact, even devising a joint model for all of the marks would pose
a challenge (e.g., modeling gender as a function of location, age, race, education, and cause of death).


The framework is general and can be adapted to a variety of settings.  Here, we presented two types of univariate responses---a
normal distribution and a truncated Poisson distribution---but any geostatistical model can be used, including models for multivariate outcomes.
 For instance, we presented a normal model with constant variance over space, but one can instead
use models with spatially-varying variances.  One such example would be the
modeling of individuals' income, where individuals in rural areas may have less variable
 incomes than their urban counterparts.
While we used (modified) predictive processes to achieve dimension reduction in the modeling of the
intensity and response surfaces, one may use other dimension reduction approaches such as those described in \citet{wikle:hand} if they are more appropriate for a given setting.

Although the methodology is intended to generate {fully} synthetic datasets, agencies can easily adapt the procedures to generate partially synthetic data.  A natural option would be
%
to reorganize the modeling strategy such that sensitive attributes are modeled conditionally on the non-sensitive data.
For instance, it may be the case that only the individuals' locations are deemed sensitive; in this case, we could define
the intensity surfaces in Section~\ref{sec:methods} as functions of the non-categorical marks as well as the categorical marks.
While intuitive structures for modeling these intensity surfaces can be difficult to define, the end result would be similar to
the CART-based approach used by \citet{wangreiter}.
}

These methods could naturally be extended to the spatio-temporal setting.  In its simplest form, one could treat time as discrete (e.g., weekly incidents of a communicable disease) and consider each time-point as a separate categorical mark.  The case of continuous time is more challenging, both in the modeling stage and in the generation of synthetic data.  While one could treat space-time as a 3-dimensional object, generating synthetic space-time coordinates could easily require evaluating the 3-dimensional ``intensity object'' at over 100,000 different space-time proposal coordinates.  For our example, we used a $50\times50$ grid of spatial locations; expanding this to include 50 time-points suddenly requires 125,000 unique space-time coordinates.  This is a subject of future research.

Another area for future work is to provide a comprehensive investigation of what to do when disclosure risks are deemed unacceptably high.
For example, inferential disclosure risks may be unacceptably high when the models overfit the
 data or when the confidential data consist of relatively homogeneous clusters of individuals.  In these cases, we speculate that
agencies may need to incorporate differential amounts of spatial smoothing for locations and cases at high risk.  
Obviously, the challenge here will be to further reduce disclosure risks without sacrificing data utility.

\section*{Acknowledgements}
        This research was partially supported by the U.S. National Science Foundation (NSF) and the U.S. Census Bureau under NSF grants SES-1132031 and SES-1131897, funded through the NSF-Census Research Network (NCRN) program.

\bibliographystyle{jasa}
\bibliography{mpp}

\end{document}